\documentclass[12pt,twoside]{article}

\begin{document}\hbadness=10000\thispagestyle{empty}
%%%%%%%%%%%%%%%%%%%%%%%%%%%%%%%%%%%%%%%%%%%%%%%%%%%%%%%
\pagestyle{myheadings}
\title{{\bf On supercorrelated systems and phase space entrainment}}    
\author{$\ $\\
{\bf H.-T. Elze and T. Kodama} 
\\ $\ $\\ 
Instituto de F\'{\i}sica, Universidade Federal do Rio de Janeiro \\ 
C.P. 68.528, 21941-972 Rio de Janeiro, RJ, Brazil\footnote{E-mail: thomas@if.ufrj.br, 
tkodama@if.ufrj.br}
}
\vskip 0.5cm
\date{December 2004 (revised)}% Deleting this produces today's date 
\maketitle
\vspace{-8.5cm}
%{\bf preprint--XX}
\vspace*{8.0cm}
%%%%%%%%%%%%%%%%%%%%%%%%%%%%%%%%%%%%%%%%%%%%%%%%%%%%%%%%%%%%%%%
\begin{abstract}{\noindent
It is demonstrated that power-laws which are modified by logarithmic 
corrections arise in supercorrelated systems. 
Their characteristic feature is the energy attributed to a state 
(or value of a general cost function) which depends nonlinearly on the 
phase space distribution of the constituents. A onedimensional dissipative 
deterministic model is introduced which is attracted to a supercorrelated 
state (phase space entrainment). Extensions of this particular model may have 
applications in the study of transport and equilibration phenomena, 
particularly for supply and information networks, or for 
chemical and biological nonequilibrium systems, while the qualitative 
arguments presented here are believed to be of more general interest. 
\vskip 0.2cm
\noindent
%PACS:
}\end{abstract}
%%%%%%%%%%%%%%%%%%%%%%

%%%%%%%%%%%%%%%%%%%%%%%
\section{Introduction}
%%%%%%%%%%%%%%%%%%%%%%%
Power-laws are omnipresent in natural or man-made systems \cite{Zipf,Mandelbrot}. 
They arise in the areas of high-energy particle physics, condensed matter, complexity, 
sociology, and linguistics, to name a few. They are an important  
feature of Tsallis statistics \cite{Tsallis,TsallisRev}. While numerous 
out-of-equilibrium statistical systems are known showing this behavior, 
only rarely the dynamics is understood that leads to it.  

In this letter we study  
systems that are correlation dominated, in a sense to 
be more accurately defined. We show that in this class of models  
the asymptotic spectra are essentially given by power-laws, 
however, modified by logarithmic corrections. They appear 
quite similar from a phenomenological point of view. 
 
Recently, for example, 
most interesting transient dynamical behavior has been observed in relativistic 
heavy-ion collisions \cite{QM}. It is observed that apparently thermodynamics 
rules, with a local equation of state, in particular, and 
commencing at unexpectedly early times 
(of order $0.3-0.9\mbox{x}10^{-24}\mbox{s}$) 
during the collision process. However, the experiments also indicate that typical 
one-particle observables are not correspondingly thermalized by that time. 
Theoretical investigations based on quantum field theoretical models  
support the view that such `prethermalization' is indeed produced \cite{Berges}.  
This is accompanied by spectra, which show considerable deviations from 
Fermi-Dirac, Bose-Einstein, or Boltzmann distributions, with uniform parameters,     
and which relax to the equilibrium form only on a much longer time scale. 

However, a heuristic picture which allows us to understand 
essential features of the underlying processes generally has been lacking. 
Recently, we have shown that such behavior can be interpreted in terms of certain 
correlations generated dynamically and, in fact, power-laws can emerge as a 
consequence \cite{Us}. 
Presently, we start with a different class of models. In some limits analytical 
results can be obtained, the models can be easily generalized, and may be applied in quite different   
contexts. We demonstrate the emergence of 
log-modified power-laws as a robust feature in {\it supercorrelated systems}   
where correlations among the distributions of the constituents govern the statistical 
behavior. 

Other recent works invariably invoke stochastic processes 
in the explanation of power-laws, e.g., in Refs.\,\cite{Us,Denisov,Beck,JR,Biro}. 
In distinction, we present a deterministic dissipative model where microscopic stochastic 
processes are subsumed in suitable (anti)damping terms of otherwise Hamiltonian equations 
of motion. We demonstrate that {\it phase space entrainment} leads to supercorrelation 
here and, thus, ultimately to the log-modified power-laws. 

%%%%%%%%%%%%%%%%%%%%%%%%%%%%%%%%%%
\section{Supercorrelated systems}
%%%%%%%%%%%%%%%%%%%%%%%%%%%%%%%%%%
Specifically, we consider 
a two-dimensional `phase space' spanned by a discrete `spatial' coordinate $x\in\{x_i,\;i=1,\dots ,L\}$  
(with periodic boundary condition)  
together with a discrete `energy' coordinate $E\in\{E_j,\;j=1,\dots M\}$. We would like to understand the 
statistical behavior of a system consisting of $L$ `particles', which are distributed randomly over the 
energy $E$, however, with exactly one particle per lattice site $x$.   

The interpretation assigned to the coordinates may vary according to circumstances. 
In particular, the `energy' may generally represent any {\it cost function}. 

Crucially, we assume that there is a characteristic energy $E_j$ associated with the $j$-th state of the 
system. Furthermore, it is the product of this energy times a  
correlation function $C_j$ which yields the relevant effective energy. Only the latter will 
determine the relative weight of terms in a  
sum over states. More precisely, the correlation function $C_j[p]$ is determined by a product of particle 
distributions $p$. 
The number of factors of $p$ involved will be called the correlation degree $c$. 
(Out of a sum of terms of different 
degree only the one with smallest $c$ matters asymptotically.) 

Before we proceed, we define this correlation energy in various examples: 
\begin{itemize}
\item 
If the correlation is between a particle at site $x$  
and a second one 
a distance $\Delta x$ away, then the contribution 
to the total energy is: 
\begin{equation}\label{correlation1}
\sum_jE_jC_j[p]\equiv\sum_{i,j}E_jp(x_i,E_j)p(x_i+\Delta x,E_j)
\;\;. \end{equation}
Note that all possible pairs ($c=2$) are counted here exactly once, due to the periodic boundary condition.
\item
Another case of interest is that the correlation of the particles does not depend on 
the mutual distance, corresponding to the limit that the interaction length is larger than 
the system size, $\Delta x>x_L-x_1$:  
\begin{equation}\label{correlation3}
\sum_jE_jC_j[p]\equiv\sum_{i_1<i_2<\dots <i_c;j}E_j
p(x_{i_1},E_j)p(x_{i_2},E_j)\cdot\dots\cdot p(x_{i_c},E_j)
\;\;, \end{equation}
with $c$ ordered factors of $p$, not restricted otherwise because of lattice periodicity.
\item 
Finally, we may have an anticorrelation instead of the previous case: 
\begin{equation}\label{correlation4}  
\sum_jE_jC_j[p]\equiv\sum_{i_1;j}E_jp(x_{i_1},E_j)\Big (1-
\sum_{i_1<i_2<\dots <i_c}p(x_{i_2},E_j)\cdot\dots\cdot p(x_{i_c},E_j)\Big )
\;\;. \end{equation}
Numerous further variants can be imagined.  
\end{itemize}  

It seems worth while to discuss various features here.  
First of all, it is important to realize that 
the energies $E_j$ need neither be related to single-particle nor to bound-state energy levels.   
The energy $E_j$, or a fraction thereof, is contributing here to the extent that the particle 
distribution maximizes the correlation function, or minimizes it, depending on the details of $C_j[p]$.
Thus, we are studying the nonlinear problem of statistical distributions of particles (one per lattice site) 
over states $j$ when  
the effective energy depends itself on the distribution, $E_j\rightarrow E_jC_j[p]$; generally, 
it may as well depend on $j$ explicitly. 
Hence, we call such systems {\it supercorrelated}. 

The relation between supercorrelated statistics 
and dissipative yet deterministic dynamics shall be discussed in the following section of this letter.  

Furthermore, a little thought shows that the examples of Eqs.\,(\ref{correlation1}) and  (\ref{correlation3}) will tend 
to enhance the probability of rare configurations of the system, relative to the case of Boltzmann-Gibbs 
equilibrium statistics. 
Similarly, the example (\ref{correlation4}) will enhance the frequent ones. These 
expectations will be confirmed by the following calculations.   

We look for the least biased probability distribution $p(x,E)$ describing an ensemble of 
such systems, subject to the additional constraints representing the average total energy $\langle E\rangle$ and the fixed particle number $L$. Optimizing 
the {\it Shannon (information) entropy} functional, we are lead to a  
variational determination of the probability distribution: 
\begin{eqnarray} 
&\;&\delta\Big (\sum_{i,j}p(x_i,E_j)\ln [p(x_i,E_j)]+\beta (\sum_jC_j[p]E_j-\langle E\rangle )
\nonumber \\ [1ex] \label{variation}
&\;&\;\;\;\;
+\sum_i(\lambda (x_i)-1)(\sum_jp(x_i,E_j)-1)\Big )\;=\;0
\;\;. \end{eqnarray}
Here $\beta$ and $\lambda$ are Lagrange multipliers incorporating the average energy 
and fixed particle number constraints. 

A remark is in order here. The mathematical foundations of statistical mechanics of supercorrelated 
systems, as introduced in this work, certainly need further study. Presently, we do not presume a  
detailed relation to thermodynamics, even though some of the following results are 
suggestive in this direction
(see also the discussion in Ref.\,\cite{Us}). However, the application of the variational principle 
for the Shannon entropy 
is justified, since the searched-for (normalized) distribution $p(x,E)$ refers to a closed system 
which is thought to be realizable in a sufficiently large number of cases, thus defining the ensemble 
\cite{Haken,Schuster}. What is nonstandard here, is the nonlinear (in $p$) constraint incorporating 
the correlation energy or cost function.\footnote{Restriction to any particular form of the constraints 
is not required by the arguments leading to the variational principle for the 
Shannon entropy \cite{Haken,Schuster}.} -- Since the present nonlinearities bear some resemblance 
to the ones of escort distributions, we anticipate that further developments will allow to relate the 
present approach to superstatistics \cite{BC}, where such distributions can be successfully 
dealt with in a canonical framework \cite{B}.

We proceed to perform the variations and, thus, obtain 
the following set of equations: 
\begin{eqnarray}\label{normalization} 
&\;&\sum_jp(x_i,E_j)=1
\;\;, \\ [1ex] \label{energy}
&\;&\sum_jC_j[p]E_j
=\langle E\rangle 
\;\;, \\ [1ex] \label{distribution}
&\;&\ln [p(x_k,E_j)]+\frac{\beta E_j}{p(x_k,E_j)}
\sum_{i_1<\dots <i_c}'
p(x_{i_1},E_j)\cdot\dots\cdot p(x_{i_c},E_j)
\;+\lambda (x_k)=0
\;\;, \end{eqnarray} 
where we chose the example of Eq.\,(\ref{correlation3}) in the 
last equation; the prime on the sum indicates that the sum is only over  
those terms which contain the factor $p(x_k,E_j)$. 

Combining Eqs.\,(\ref{normalization}) and (\ref{distribution}), the Lagrange multiplier 
$\lambda$ can be eliminated to obtain implicitly the normalized distribution: 
\begin{equation}\label{normdistr}
p(x_k,E_j)=Z^{-1}(x_k)\exp\Big [-\frac{\beta E_j}{p(x_k,E_j)}
\sum_{i_1<\dots <i_c}'
p(x_{i_1},E_j)\cdot\dots\cdot p(x_{i_c},E_j)
\Big ] 
\;\;, \end{equation} 
with the partition function, 
\begin{equation}\label{Z} 
Z(x_k)\equiv\sum_j\exp\Big [-\frac{\beta E_j}{p(x_k,E_j)}
\sum_{i_1<\dots <i_c}'
p(x_{i_1},E_j)\cdot\dots\cdot p(x_{i_c},E_j)
\Big ]
\;\;. \end{equation}
In principle, the Eq.\,(\ref{energy}) serves to fix $\beta$. However, instead, 
we may choose to work with the distribution as it is, considering $\beta$ 
as a macroscopic parameter characterizing the ensemble.\footnote{This is analogous 
to working with a standard canonical ensemble, where $\beta$ would correspond to the inverse temperature, $\beta\equiv 1/T$, in units where $k_{\mbox{B}}=1$.}   

In the absence of localized impurities or dissipative effects, the system cannot spontaneously break 
(lattice) {\it translation invariance}. With this simplification, the distribution 
has to be homogeneous in space, 
\begin{equation}\label{homogeneity} 
p(x_k,E_j)=p_x(E_j) 
\;\;, \end{equation}  
i.e., the same at any chosen lattice site. This 
allows us to simplify Eq.\,(\ref{normdistr}): 
\begin{equation}\label{normdistr1}
p_x(E_j)=Z_x^{-1}\exp\Big [-\beta N_cE_jp_x^{c-1}(E_j)\Big ] 
\;\;, \end{equation} 
where $Z_x$ is the correspondingly simplified partition function, cf. Eq.\,(\ref{Z}). The combinatorial factor is 
given by: 
\begin{equation}\label{Nc} 
N_c\equiv c\left (
\begin{array}{c} L \\ c \end{array}
\right ) 
\;\;. \end{equation}
Note that for $c=1$ we recover Boltzmann statistics, however, with a renormalized temperature, 
which is due to the presence of $L$ identical copies of the system, in this case. 
  
The transcendental Eq.\,(\ref{normdistr1}) is of the form $P=\exp (-\eta P)$, with 
$P\equiv (p_xZ_x)^{c-1}$ and $\eta\equiv (c-1)\beta N_cE_jZ_x^{1-c}$. 
Its solution can be given in terms of the Lambert $W$-function, $P(\eta )=W(\eta )/\eta $. 
In the limit  
$\eta\ll 1$, one obtains the following approximation:  
\begin{equation}\label{normdistr2} 
p_x(E_j)=Z_x^{-1}\frac{1}{\Big (1+(c-1)Z_x^{1-c}\beta N_cE_j\Big )^{1/(c-1)}}
\;\;,\;\;\;\eta\ll 1
\;\;, \end{equation}   
which yields a {\it power-law} with exponent dictated by the correlation degree. 
This is a typical form of the Tsallis distribution \cite{Tsallis,TsallisRev}, or of Zipf's law as 
generalized by Mandelbrot \cite{Zipf,Mandelbrot}. 
In the opposite limit, a {\it log-modified power-law} results: 
\begin{equation}\label{normdistr22} 
p_x(E_j)=Z_x^{-1}\left (\frac{F\Big ((c-1)Z_x^{1-c}\beta N_cE_j\Big )}
{(c-1)Z_x^{1-c}\beta N_cE_j}\right )^{1/(c-1)}
\;\;,\;\;\;\eta\gg 1
\;\;, \end{equation}   
with the function $F$ defined by: 
\begin{equation}\label{F}
F(x)\equiv \ln (x)\Big (1+\ln (\ln (x))/\ln (x)\Big )^{-1} 
\;\;. \end{equation}  
In the example of Eq.\,(\ref{correlation1}), we would 
find similar results in the homogeneous limit, differing only by the appropriate 
correlation degree and the combinatorial factor. 

We remark that there is a considerable similarity between the Zipf-Mandelbrot law (\ref{normdistr2}) and the 
distribution (\ref{normdistr22}), when the former is also applied in the limit $\eta\gg 1$. 
In a log-log representation one could perhaps be mistaken for the other (for a limited 
range of data) by readjusting the parameters.    

In any case, having obtained the probability distribution $p_x(E)$ for positive integer values of $c$, 
it is tempting to extend the definition of the correlation degree to all real positive 
numbers. In Eq.\,(\ref{Nc}) this is implemented by employing the Gamma function instead 
of factorials, and it can obviously be done in Eq.\,(\ref{normdistr1}),  
for $c>1$; a comment concerning $c<1$ will follow shortly.  
However, can we understand such analytic continuation also in terms of 
the picture of supercorrelated systems? 
  
Physically, we expect that supercorrelations which are not fully developed somehow 
should correspond to noninteger correlation degree here. 
Indeed, 
let us focus on situations with $c$ sufficiently close to an integer. Here, an expansion  
provides a hint for the interpretation. 
For example, we may consider the phenomenologically interesting case $c=1+\gamma$, 
with $\gamma$ sufficiently small \cite{TsallisRev}. We obtain: 
\begin{equation}\label{pcsmall}
p^{1+\gamma}=p\cdot\left 
(1-\gamma q-\frac{1}{2!}\gamma (1-\gamma )q^2+\mbox{O}(q^3)\right )
\;\;. \end{equation}  
%Of course, one would like to insert $q=1-p$ on the right-hand here. 
%However, doing this and rearranging the series in powers of $p$ 
%leads to an asymptotic series, the coefficients of which diverge essentially after the first 
%one (which vanishes). 
We recognize a particular form of anticorrelation. For example, this can correspond to a term 
$\propto p(x_i)[1-\gamma q(x_{i}+\Delta x)]$, which presents a correlation 
with a `hole' distribution, $q\equiv 1-p$, modifying the leading Boltzmann term.  
This should be compared to the different correlation and anticorrelation  
defined in Eqs.\,(\ref{correlation1}) and (\ref{correlation4}), respectively.       

To complement these remarks, we 
briefly recall the anticorrelation of Eq.\,(\ref{correlation4}), instead of Eq.\,(\ref{correlation3}). 
The result replacing Eq.\,(\ref{normdistr1}) is simply:     
\begin{equation}\label{antidistr}
p_x(E_j)=Z_x^{-1}\exp\Big [-\beta E_j\left (1-N_cp_x^{c-1}(E_j)\right )\Big ] 
\;\;. \end{equation} 
Here, for $c>1$, the second term in the exponent presents a small 
correction to the leading Boltzmann term, which vanishes exponentially for $\beta E_j\gg 1$.  
  
An interesting situation arises for $c<1$. -- Both, the Eqs.\,(\ref{normdistr1}) and (\ref{antidistr}) 
cease to have solutions for $p_x(E_j)$ for sufficiently large $\beta E_j$ in 
this case. This is reflected in the pole which arises in the power-law result of Eq.\,(\ref{normdistr2}), when continued to $c<1$. Similarly, for example, the power-law obtained 
here from Eq.\,(\ref{antidistr}), 
\begin{equation}\label{antidistr1}
p_x(E_j)=Z_x^{-1}\Bigg (\frac{1+(1-c)Z_x^{1-c}\beta N_cE_j}{1+(1-c)\beta E_j}\Bigg )^{1/(1-c)} 
\;\;, \end{equation} 
breaks down, when unavoidably $p_x(E_j)\rightarrow 1$, for sufficiently large energy. 
Therefore, a cut-off on $E_j$ is necessary here, as discussed in the 
applications of Tsallis statistics before \cite{TsallisRev}. 

%%%%%%%%%%%%%%%%%%%%%%%%%%%%%%%%%%%%%%%%%%%%%%%%%%%%%%%%%%
\section{Via phase space entrainment to supercorrelation}
%%%%%%%%%%%%%%%%%%%%%%%%%%%%%%%%%%%%%%%%%%%%%%%%%%%%%%%%%%
It is {\it not} the purpose of this section to develop the statistical mechanics of 
supercorrelated systems from the bottom up. Since the present approach is entirely new, 
our modest aim here is to show in a simple example, based on analyzing the 
asymptotic solutions of the equations of motion, how {\it phase space entrainment} 
occurs and leads to {\it supercorrelations} as defined in the previous section. --  
An extension of the simplest two-body system to a more realistic 
one-dimensional lattice model and its mean field analysis are contained in a first  
version of this letter \cite{v1}. 

Therefore, fundamental questions, such as 
related to the (non)thermalization of the Hamiltonian system presented in the following, 
to the H-theorem, or to the fluctuation-dissipation theorem, when the system is made dissipative 
in a particular way by incorporating phaenomenological (anti)damping terms in the 
equations of motion, must be deferred to future microscopic statistical mechanics 
studies of such systems. 

Here, to begin with, correlation energies, such as defined in Eqs.\,(\ref{correlation1})--(\ref{correlation4}),  
must reflect an underlying microscopic dynamics in a coarse-grained way. In order to further illustrate the 
notion of correlation energy, we consider here a {\it two-body Hamiltonian}, 
${\cal H}(\Pi_1,E_1;\Pi_2,E_2)$, describing the interaction of two `cells' (called 
`particles' before) localized at two given sites ``1'' and ``2'' of, for example, the  previous onedimensional lattice. 
In distinction to 
ordinary Hamiltonian dynamics, we treat the {\it local energies}, $E_{1,2}$, in place of a  
coordinate, together with the canonical momenta, $\Pi_{1,2}$. Phase space consists  
of the 2x2-dimensional $(\vec\Pi,\vec E)$-space ($E_{1,2}>0$). 

The precise form of our Hamiltonian is solely dictated by mathematical convenience 
in modelling the following phaenomenological features: 
\begin{itemize} 
\item two subunits (cells, agents, etc.) of the system, i.e. sites ``1'' and ``2'', 
exchange energy (adjust a potential difference across a biological membrane, trade a 
commodity, etc.) in an oscillatory way; 
\item the total amount of energy (base potential, commodity, etc.) distributed in the system 
is conserved; 
\item the ensuing Hamiltonian dynamics may be nonlinear; 
\item the exchange process is damped, while preserving the overall conservation of 
energy (etc.). 
\end{itemize}  
We will continue to use the term `energy' henceforth. However, the versatility of such a model 
should be kept in mind. We believe that this type of systems is generic in that it produces 
some kind of supercorrelation. 

Employing more convenient continuum notation,  
the {\it average energy} $\langle {\cal E}\rangle$ corresponding to ${\cal H}$ 
for a pair of cells is: 
\begin{equation}\label{Hamiltonian} 
\langle {\cal E}\rangle =(2\pi )^{-2}
\int\mbox{d}\Pi_1\mbox{d}E_1\mbox{d}\Pi_2\mbox{d}E_2
\;{\cal H}(\Pi_1,E_1;\Pi_2,E_2)p(\Pi_1,E_1;\Pi_2,E_2) 
\;\;, \end{equation} 
in terms of the phase space pair probability density $p$, and  
where our model Hamiltonian is: 
\begin{equation}\label{Hamiltonian2} 
{\cal H}(\Pi_1,E_1;\Pi_2,E_2)\equiv\frac{E_1+E_2}{\xi\zeta}
\exp [-\xi^{-2}(\Pi_1-\Pi_2)^2-\zeta^{-2}(E_1-E_2)^2]= {\cal E}
\;\;, \end{equation}
with $\xi,\zeta$ constant parameters.

Using variables $\Pi_\pm\equiv\Pi_1\pm\Pi_2$ and $E_\pm\equiv E_1\pm E_2$, and 
with the conserved total energy ${\cal E}$ of Eq.\,(\ref{Hamiltonian2}),   
the equations of motion for this Hamiltonian can be combined 
to yield: 
\begin{eqnarray}\label{Eplus}
\dot E_+&=&0
\;\;, \\ [1ex] \label{Piplus}
\dot\Pi_+&=&-2{\cal E}E_+^{\;-1}
\;\;, \\ [1ex] \label{Eminus}
\dot E_-&=&-\frac{4{\cal E}}{\xi^2}\Pi_-
\;\;, \\ [1ex] \label{Piminus}
\dot\Pi_-&=&\frac{4{\cal E}}{\zeta^2}E_-
\;\;, \end{eqnarray} 
or, in particular, the second order equation:  
\begin{equation}\label{oscillator}
\ddot E_-+\omega^2E_-=0\;\;,\;\;\;\omega^2\equiv
\Big (\frac{4{\cal E}}{\xi\zeta}\Big )^2
\;\;. \end{equation} 
Thus, we find a harmonic oscillator with a constant frequency. 
However, the frequency strongly depends  
on the initial conditions, being exponentially smaller when the two cells start out 
far away from each other in phase space ($\Pi_-$ or $E_-$ large), than when 
they are close. There is a rapid energy swapping in the latter case, while it 
is slow in the former. 
 
We expect internal dissipative forces in coarse-grained descriptions 
of interacting microscopic systems. Presently, this dissipation is assumed to  
conserve the total energy ${\cal E}$. Therefore, it has to change the energy sum $E_+$, at 
the expense of damping the swapping of energy between the cells, which is 
described by the oscillation of the energy difference $E_-$ (and of $\Pi_-$). This can indeed be 
implemented by modifying only two of the Eqs.\,(\ref{Eplus})--(\ref{Piminus}), 
\begin{eqnarray}\label{Eplus1}
\dot E_+&=&-2\gamma\left (E_-/\zeta\right )^2E_+ 
\;\;, \\ [1ex] \label{Eminus1}
\dot E_-&=&-\frac{4{\cal E}}{\xi^2}\Pi_--\gamma E_-
\;\;, \end{eqnarray} 
keeping the other two. This results in the damped oscillator equation: 
$\ddot E_-+\gamma\dot E_-+\omega^2E_-=0$, where $\gamma$ is the damping constant,  
and with $\omega^2$ the same as before.   

The following properties can be read off immediately from 
the equations of motion (\ref{Piplus}), (\ref{Piminus}), (\ref{Eplus1}), and (\ref{Eminus1}): 
({\bf A}) The amplitude of the oscillating energy difference $E_-$ (and its derivative) is 
damped to zero ($\propto\exp [-\gamma t/2]$). -- 
({\bf B}) The energy sum $E_+$ for the two cells decreases monotonically from its initial value  
and saturates exponentially (with rate $\gamma$) at its finite asymptotic value. -- 
({\bf C}) The oscillating `momentum' difference $\Pi_-$ decreases 
exponentially to zero ($\propto\exp [-\gamma t/2]$). -- 
({\bf D}) The corresponding sum $\Pi_+$ decreases monotonically ($\propto -t$) for late times. 
Note that $\Pi_+$ does not enter the Hamiltonian.

From these observations we obtain the 
asymptotic time dependence of each cell's `coordinate' and `momentum':  
\begin{eqnarray}\label{E12}
E_{1,2}(t)&=&\frac{1}{2}(E_+\pm E_-)\;\approx\; a_+(1+ae^{-\gamma t})\pm a_-e^{-\gamma t}
\;\;, \\ [1ex] \label{Pi12}
\Pi_{1,2}(t)&=&\frac{1}{2}(\Pi_+\pm\Pi_-)\;\approx\; b_+(b-t-b'e^{-\gamma t})\pm b_-e^{-\gamma t}
\;\;, \end{eqnarray}
with constants depending on the initial conditions, besides $\xi,\zeta$, and $\gamma$ and $\omega$ 
(``1'' with upper signs, ``2'' with lower ones). 
We suppress oscillating factors here, thus emphasizing the presently 
important terms; using the full solutions, it can easily be verified that they do 
not affect the following argument. 

The dynamics described by Eqs.\,(\ref{Piplus}), (\ref{Piminus}), (\ref{Eplus1}), and 
(\ref{Eminus1}) is dissipative. Obviously, 
there is a onedimensional attractor, see Eqs.\,(\ref{E12})--(\ref{Pi12}). However, it is 
the pairing of variables which is most characteristic: the two cells considered 
are forced to follow similar trajectories in phase space, which converge asymptotically 
to one and the same (even with oscillation factors included). 
Simultaneously, the Hamiltonian is a constant of motion, 
${\cal H=E}$, i.e., the overall energy (or other quantity represented by the cost 
function) is conserved. 

This {\it phase space entrainment} is an essential ingredient for 
supercorrelated statistics in this illustrating example. 
Let us consider an ensemble of such pairs of cells with distributed initial 
conditions. Then, the Liouville equation, including the appropriate dissipative terms 
($\propto\gamma$), describes the evolution of the ensemble in phase space. {\it Any} 
solution of this equation for the probability density can be written in the form: 
\begin{equation}\label{Liouvillesol} 
p(\vec\Pi ,\vec E;t)\equiv p\left (\vec\Pi-\vec\Pi (t),\vec E-\vec E(t)\right ) 
\;\;, \end{equation}   
where the vectors collect the components ``1,2'', and where $\vec\Pi (t),\vec E(t)$ 
present a solution of the equations of motion, e.g., asymptotically 
as in Eqs.\,(\ref{E12})--(\ref{Pi12}). 

Considering initial distributions which 
are factorized, akin to Boltzmann's `Stosszahlansatz' (molecular chaos),  
$p(\vec\Pi ,\vec E;t)=p\left (\Pi_1-\Pi_1(0),E_1-E_1(0)\right )
\cdot p\left (\Pi_2-\Pi_2(0),E_2-E_2(0)\right )$, one obtains correspondingly 
factorized solutions at all times. 

Employing this fact, we  
further evaluate Eq.\,(\ref{Hamiltonian}) together with (\ref{Hamiltonian2}). 
Since the Hamiltonian is a constant of motion, the average 
energy can be evaluated at any (late) time. 
Neglecting exponentially small corrections, see Eqs.\,(\ref{E12})--(\ref{Pi12}), 
we obtain: 
\begin{eqnarray}  
\langle {\cal E}\rangle &=&
\int\frac{\mbox{d}\Delta_\Pi\mbox{d}\Delta_E}{2\pi}\; 
\frac{2}{\xi\zeta}\exp [-\xi^{-2}\Delta_\Pi^2-\zeta^{-2}\Delta_E^2]
\int\frac{\mbox{d}\Pi\mbox{d}E}{2\pi}\; 
(E+\Delta_E/2)
\nonumber \\ [1ex] \label{Eaverage}
&\;&\;\;\cdot\;p\Big (\Pi -b_+(b-t),E-a_+\Big )
        \cdot\;p\Big (\Pi -b_+(b-t)+\Delta_\Pi ,E-a_++\Delta_E\Big )
\\ [1ex] \label{Eaverage1} 
&\approx&\int\frac{\mbox{d}\Pi\mbox{d}E}{2\pi}\; 
E\;p(\Pi,E)\;p(\Pi,E)
\;\;, \end{eqnarray} 
where we substituted $\Pi_2\equiv\Pi +\Delta_\Pi$ and $E_2\equiv E+\Delta_E$, and with 
$\Pi,E$ instead of $\Pi_1,E_1$ earlier; the further approximation here consists in 
replacing the Gaussians by $\delta$-functions (for sufficiently small $\xi ,\zeta$), 
where, more generally, gradient corrections come into play. The constant shift 
of the energy by $-a_+$ has been absorbed into a redefinition of the distribution $p$, 
i.e., of the initial condition. Thus, we obtain 
a phase space generalization of the simplest case of {\it supercorrelation} ($c=2$) and 
corresponding correlation energy, 
cf. Eq.\,(\ref{correlation1}), in our two-cell example.

We remark that the intermediate result of Eq.\,(\ref{Eaverage}) indicates a  
natural generalization of the correlation energies of 
Eqs.\,(\ref{correlation1})--(\ref{correlation4}), namely allowing a certain 
spread of the energy when summing over states.   

Having demonstrated the emergence of supercorrelations in an almost trivial example, 
which in turn form the starting point of the derivations of the previous section, 
we conclude here with two remarks concerning more realistic generalizations of this 
model. 

A many-body system made up from subunits which interact pairwise, as investigated here, 
is bound to have rich transient dynamics, due to locally 
varying and time dependent effective oscillator frequencies \cite{v1}. In the two-cell 
example, the one frequency depends strongly on the initial condition. However, energy 
conservation prevents its variation. In general, local energy conservation will amount 
to only a global constraint and the frequencies vary locally, depending on the system  
variables. This may initially lead to the formation of `hot spots',  
with rapid energy swapping between neighbouring cells, and relatively inactive 
regions elsewhere. 

Furthermore, without adding further couplings to the basic Hamiltonian, the model 
{\it cannot thermalize}, i.e., sustain a certain amount of fluctuations. 
This can easily be changed in a future extension by modifying the decription 
of damping of the relative motion between interacting cells and feedback of 
the dissipated energy into the system. 
For example, a damping term like for the van\,der\,Pol oscillator could be 
considered. This would raise the dimensionality of the underlying 
attractor and generate more interesting behavior, especially in a many-body 
system. 

%%%%%%%%%%%%%%%%%%%%%%
\section{Conclusions}
%%%%%%%%%%%%%%%%%%%%%%
This letter has addressed the dynamical origin of power-laws. 
We have introduced the notion of {\it supercorrelated 
systems} where, as a result of interactions among its constituents, 
the energy or a more general cost function depend effectively on the 
distribution of the constituents. 

This leads to a nonlinearity in the evaluation of sums over states, 
such as calculating the partition function, and in the determination 
of the ``would-be-equilibrium'' distributions. We showed that asymptotically 
the resulting distributions are (log-modified) power-laws, with parameters 
determined 
by the correlation degree characterising the correlation energy, 
see Eqs.\,(\ref{correlation1})--(\ref{correlation4}), besides macroscopic 
parameters of the system. 
  
These results follow from a simple statistical analysis and indicate 
that a wide range of models can be conceived which lead to such distributions. 
The crucial point consists in the identification of the relevant correlation 
energy, depending on the underlying dynamical model. 
    
We presented a very simple deterministic dissipative model of interacting 
cells, where the interaction causes redistribution of the energy content 
among cells. While the overall energy of the system is conserved, the 
rate of change of gradients of the energy content of cells is  
damped by a dissipative force. 

The model has interesting nonlinear features which 
deserve further study. Presently, we have considered only two interacting subunits 
of the system, while also a mean field approximation of a onedimensional lattice model 
has been studied with analogous results \cite{v1}. In its present form, the model cannot 
lead to thermodynamic 
equilibrium. However, an extension achieving this can easily be defined. 
For our purposes most important has been the fact that the dissipation leads 
to {\it phase space entrainment}, i.e. the oscillator degrees of freedom  
are forced to converge on a simple attractor in phase space. 
  
While the effectively linear dynamics of the present model is rather trivial, it 
demonstrates how supercorrelations arise, which in turn generate (log-modified) power-laws.  
  
It will be very interesting to learn about the time dependent behavior of 
the system when nonlinearities and further thermalizing 
couplings are introduced. Then, interesting questions may be raised concerning the (non)equilibrium 
character of the obtained distributions and the thermodynamic interpretation of 
the information entropy, on which our derivations are based.   
Thus, one may find a 
testing ground for ideas about the prethermalization mentioned 
in the Introduction. 
Our model presents a coarse-grained picture of microscopic dynamics, 
which may be difficult to access directly in complex systems.
Thus, refined versions may have interesting applications in 
studies of transport and equilibration properties of condensed matter, 
polymer, or generally network structures.   

%%%%%%%%%%%%%%%%%%%%%%%%%%%%%%
\subsection*{Acknowledgements} 
%%%%%%%%%%%%%%%%%%%%%%%%%%%%%%
We wish to thank C.E.\,Aguiar and T.\,Koide for discussions. One of us (HTE) wishes to 
thank P.\,Quarati and A.\,Lavagno for kind hospitality at Politecnico di Torino and 
for the interesting discussions, which were helpful in revising this paper. The anonymous 
referee is warmly thanked for his constructive criticism. -- 
This work has been supported in part by CNPq, FINEP, FAPERJ, and CAPES/PROBRAL.

%%%%%%%%%%%%%%

\end{document}